\UseRawInputEncoding
\documentclass[manuscript]{acmart}

\AtBeginDocument{%
  \providecommand\BibTeX{{%
    \normalfont B\kern-0.5em{\scshape i\kern-0.25em b}\kern-0.8em\TeX}}}

\setcopyright{acmcopyright}
\copyrightyear{2022}
\acmYear{2022}
\acmDOI{}

\acmConference[DIS '22]{DIS '22: ACM SIGCHI Conference on Designing Interactive Systems}{June 13--June 17, 2022}{online}
\acmBooktitle{DIS '22: ACM SIGCHI Conference on Designing Interactive Systems (DIS),
  June 13--June 17, 2022, Online}
\acmPrice{15.00}
\acmISBN{978-1-4503-XXXX-X/18/06}

\usepackage{url}
\usepackage{enumerate}

\def\markup{0}  
\if\markup 1
\usepackage{soul}
\newcommand{\rv}[1]{{\leavevmode\color{blue}#1}}
\else
\newcommand{\rv}[1]{#1}
\newcommand{\st}[1]{}
\fi

\begin{document}










\title[Understanding Blind or Low Vision Streamers' Perceptions of Content Curation Algorithms]{``It Feels Like Being Locked in A Cage'': Understanding Blind or Low Vision Streamers' Perceptions of Content Curation Algorithms}

\author{Ethan Z. Rong}
\affiliation{%
  \institution{Computational Media and Arts Thrust,}
  \institution{The Hong Kong University of Science and Technology (Guangzhou)}
  \city{Guangzhou}
  \country{China}
}
\email{zrongac@connect.ust.hk}

\author{Mo Morgana Zhou}
\affiliation{%
  \institution{Department of Electrical Engineering,}
  \institution{City University of Hong Kong}
  \city{Hong Kong SAR}
  \country{China}
}
\email{mzhou25-c@my.cityu.edu.hk}

\author{Zhicong Lu}
\authornote{Corresponding Authors}
\affiliation{%
  \institution{Department of Computer Science,}
  \institution{City University of Hong Kong}
  \city{Hong Kong SAR}
  \country{China}
}
\email{zhicong.lu@cityu.edu.hk}

\author{Mingming Fan}
\authornotemark[1]
\affiliation{
  \institution{Computational Media and Arts Thrust,}
  \institution{The Hong Kong University of Science and Technology (Guangzhou)}
  \city{Guangzhou}
  \country{China}
}
\affiliation{
  \institution{Division of Integrative Systems and Design,}
  \institution{Department of Computer Science and Engineering,}
  \institution{The Hong Kong University of Science and Technology}
  \city{Hong Kong SAR}
  \country{China}
}
\email{mingmingfan@ust.hk}

\renewcommand{\shortauthors}{Rong, Zhou, Lu, and Fan}

\begin{abstract}

Blind or low vision (BLV) people were recently reported to be live streamers on the online platforms that employed content curation algorithms. Recent research uncovered algorithm biases suppressing the content created by marginalized populations. However, little is known about the effects of the algorithms adopted by live streaming platforms on BLV streamers and how they, as a marginalized population, perceive the effects of the algorithms. We interviewed BLV streamers (N=19) of Douyin --- a popular live stream platform in China --- to understand their perceptions of algorithms, perceived challenges, and mitigation strategies. Our findings show the perceived factors contributing to disadvantages under algorithmic evaluation of BLV streamers' content (e.g., issues with filming and timely interaction with viewers) and perceived algorithmic suppression (e.g., content not amplified to sighted users but suppressed within the BLV community). Their mitigation strategies (e.g., not watching other BLV streamers' shows) tended to be passive. We discuss design considerations to design a more inclusive and fair live streaming platform.

\end{abstract}

\ccsdesc[500]{Human-centered computing~Human computer interaction (HCI)}
\ccsdesc[300]{Human-centered computing~Empirical studies in HCI}

\keywords{Algorithms, Algorithmic Experience, Perceptions of Algorithms, Accessibility, Individuals with Disabilities \& Assistive Technologies, Social Media/Online Communities, Interview}



\maketitle

\begin{figure*}
    \centering
    \includegraphics[width=\textwidth]{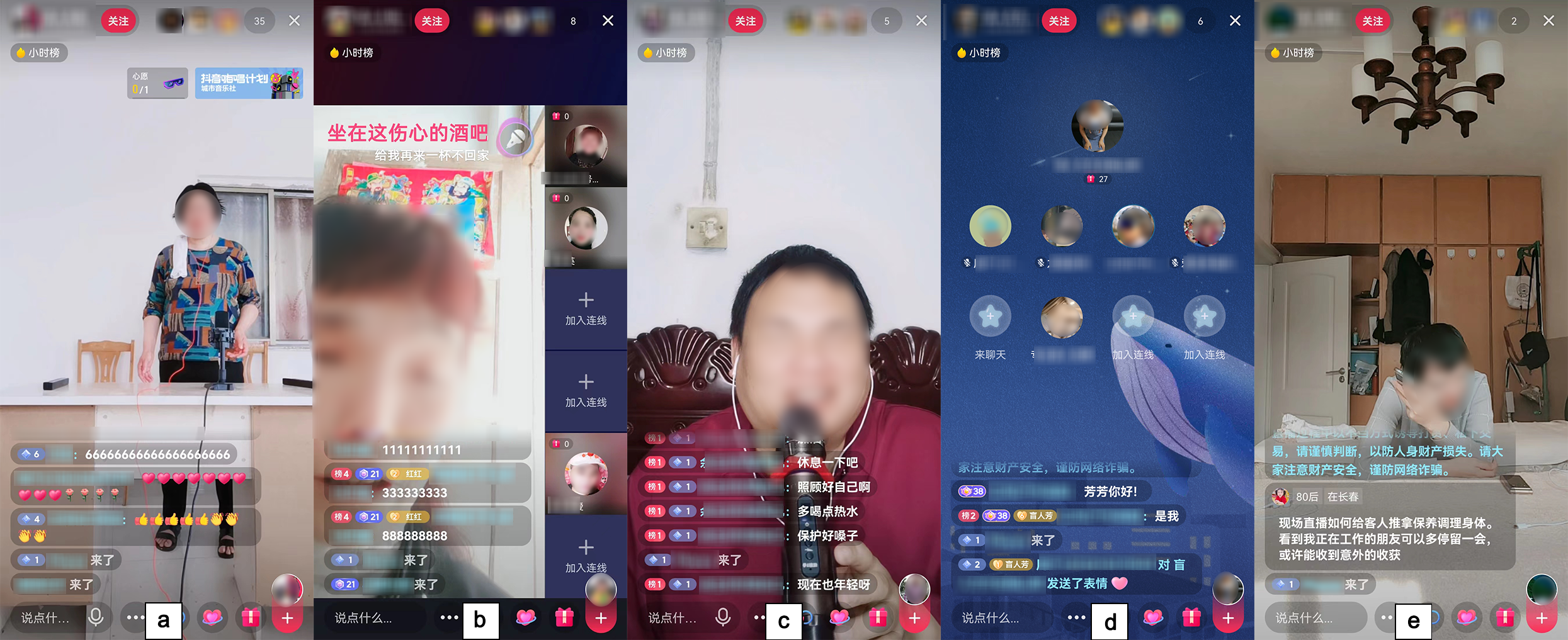}
    \Description[Screenshot of BLV streamers' live-streaming]{Five captures separately showed five live-streaming practices of BLV streamers.}
   \caption{People with visual impairments livestream varied content on Douyin, such as: (a) demonstrating how a person with visual impairments does housework (e.g., wiping a table), (b) singing, (c) answering viewers' questions relating to visual impairments, (d) chatting without turning on camera, and (e) showcasing the working environment of a massage shop.}
   \label{fig:captures}
\end{figure*}


\section{INTRODUCTION}
\label{INTRODUCTION}
The prevalence of more affordable high-speed internet and high-resolution cameras on mobile devices has enabled widespread live-streaming practices on the platforms like Tiktok, Twitch, and Youtube. Streamers share a wide range of live videos regarding pan entertainment~\cite{lu2018you}, goods selling~\cite{lu2018you,hu2020enhancing}, education content~\cite{lu2018you,chen2021towards,chen2021learning,fonseca2021knowledge,hammad2021towards}, gaming~\cite{tang2016meerkat,woodcock2019affective}, and intangible culture heritage~\cite{lu2019feel}. The ubiquity of live-streaming also attracts people with different abilities to engage, such as people with disabilities. Prior research discussed the enjoyable experiences and challenges of video game streamers with physical disabilities, ADHD (Attention Deficit Hyperactivity Disorder), and dyslexia on Twitch~\cite{johnson2019inclusion,anderson2021gamer}. \rv{Although} live-streaming is a highly visually-demanding activity (e.g., scrolling comments, filming, and filter), people who are blind or have low vision (i.e., BLV) were recently reported to be active content creators on Tiktok~\cite{alastair2021sheffield, torres2020blind,lasker2020teens,brown2020tiktok}.  As online platforms that support live-streaming (e.g., Youtube, TikTok, and Douyin) often employ algorithms to curate, select, and present contents to viewers, content creators' activities on such algorithm-driven platforms tend to affect how the algorithms curate such contents ~\cite{karizat2021algorithmic}. However, little is known \rv{about how} streamers with visual impairments perceive the effects of such algorithms and how their practices on live-streaming platforms \rv{may be shaped by} the algorithms. 

Moreover, recent work uncovered that the algorithms of video-sharing platforms were perceived to suppress the content from marginalized groups, such as people of color, the LGBT+ community, and content creators of lower socio-economic status~\cite{simpson2021you,karizat2021algorithmic}. The videos related to LGBT+ social identities were reported to be removed by Tiktok algorithms~\cite{simpson2021you}. Some anecdotes revealed that Tiktok prevented the content of disabled users (e.g., people with facial disfigurements, down's syndrome, or autism) from showing in able-bodied people's feeds~\cite{linessocial,trewin2019considerations, bozdag2013bias,robertson2019tiktok,biddle2020invisible}. \st{This raises the question of \textit{whether and to what extent BLV streamers' content is subjected to the manipulation~(e.g., removal, suppression) of the algorithm adopted by video-sharing and live-streaming platforms. How do BLV content creators perceive the algorithms? What strategies do they employ to mitigate the barriers created by the algorithms?}} \rv{This raises the questions of \textit{whether BLV content creators \textbf{perceive} there are any algorithmic challenges in relation to their status of visual impairments. If yes, what kinds of \textbf{perceived} algorithmic challenges that BLV content creators encounter? What strategies do they employ to mitigate the \textbf{perceived} barriers created by the algorithms?}}

To understand BLV content creators' perceptions of algorithms and to figure out whether they encountered similar \rv{perceived} algorithmic bias as other types of disabled people did, we must understand the challenges they encounter and the mitigation strategies they employ when interacting with the live-streaming ecosystem. 

Motivated to understand BLV content creators' perceptions and challenges of the algorithms adopted by live-streaming platforms, we conducted semi-structured interviews with BLV streamers (N=19) of Douyin (i.e., Tiktok in China).
We chose to study Douyin due to its popularity in China~\cite{lu2018you}. A survey conducted by Shenzhen Accessibility Research Association indicated that although no screen readers were perfectly compatible with Douyin, twenty-six percent of BLV respondents still kept using Douyin frequently ~\cite{zijie2019}. In Douyin, content creators not only stream but also post and share short videos. Their live-streams or videos are promoted to users' For You Page~(FYP) by Douyin's algorithms. Thus, content creators on Douyin tend to rely on algorithms to make their content visible to expected audiences heavily.


Our findings uncovered BLV content creators' perceived factors contributing to their disadvantages under the algorithmic evaluation of BLV streamers' content, such as visual effects of filming and filters, interaction with viewers, video editing, misperceptions, and trolls. Moreover, they also suffered from perceived algorithmic biases regarding content suppression. For example, participants believed that the platform did not actively promote their content to the sighted audience \rv{(e.g., participant 16 noted:“It feels like being locked in a cage, since my content can hardly reach sighted viewers.”).}  \rv{Also, they perceived that Douyin's algorithms} tended to limit the popularity of BLV streamers' content within the BLV community. BLV streamers developed mitigation strategies, such as hiding BLV identities, tagging geo-locations at downtown areas which had more potential sighted viewers, creating counter-intuitive feelings in their content, actively interacting with sighted users and providing mutual support within the BLV community. Based on the findings, we discussed the design considerations to make live-streaming platforms more accessible and inclusive for the BLV community.

In sum, we made the following contributions in this work: i) We identified BLV content creators' perceptions of how Douyin's algorithms operated in relation to BLV status; ii) We uncovered the perceived algorithmic challenges that BLV streamers encountered as well as their mitigation strategies; iii) We presented design considerations to make a more inclusive and fair live-streaming ecosystem for BLV streamers. 


\section{RELATED WORK}
\label{RELATED WORK}

We first review the prior work in practices and populations on platforms that supported live-streaming, followed by a review of algorithmic biases and stereotypes for marginalized groups.

\subsection{The Practices and Populations on Platforms that Supported Live-streaming}
\label{The Practices and Populations on Platforms that supported live-streaming}

Due to the surge of live-streaming, \st{there was} an increasing number of studies \st{investigating} investigated users' live-streaming practices on social media platforms. Prior research indicated that people from a variety of professions actively streamed online, such as professional streamers~\cite{woodcock2019affective, lu2018you} , practitioners safeguarding Intangible Cultural Heritage (ICH)~\cite{lu2019feel} , programmers~\cite{chen2021towards,hammad2021towards} , university instructors~\cite{chen2021learning,hammad2021towards} , students preparing for specific tests~\cite{wang2021study} , community leaders and reporters~\cite{dougherty2011live} . For instance, Chen et al.\cite{chen2021towards} indicated that programmers chose live-streaming to share programming knowledge and skills. Some studies also examined a wide range of live-streaming 
practices, including pan entertainment~\cite{lu2018you} , goods selling~\cite{lu2018you,hu2020enhancing} , online education~\cite{lu2018you,chen2021towards,chen2021learning,fonseca2021knowledge,hammad2021towards} , personal story sharing~\cite{lu2018you} , intangible culture promotion~\cite{lu2018you,lu2019feel}, eating~\cite{anjani2020people}, outdoor activities~(e.g. traveling, adventures)~\cite{lu2019vicariously,tang2016meerkat}, and gaming~\cite{tang2016meerkat,woodcock2019affective}. \st{Lu et al. found that streamers who promoted ICH-related content leveraged various streaming structures to showcase ICH practices, such as questions and answers, expert interviews, learners' live performances, and tutorials regarding basic knowledge.} \rv{ Lu et al.\cite{lu2019feel} found that streamers who promoted ICH-related content leveraged various streaming structures to showcase ICH practices. For example, ICH streamers held ``Questions and Answers'' sessions, interviewed experts, asked learners to perform, and delivered tutorials on basic knowledge.} Work by Lu et al.~\cite{lu2019vicariously} analyzed outdoor live-streaming in China.~\st{finding that in addition to broadcasting outdoor activities or travel in different environments~(e.g., hiking, fishing, and hunting), streamers also shared their spontaneous and unpredictable interaction with strangers (e.g., passersby on the street and the Didi~(Uber) drivers).} \rv{This work showed that in livestreams, streamers broadcasted outdoor activities or traveled in different environments (e.g., hiking, fishing, and hunting). In addition, they shared spontaneous and unpredictable interactions with strangers (e.g., passersby on streets and Didi (Uber) drivers).}

Some prior work presented that people with disabilities actively streamed. Anderson and Johnson~\cite{anderson2021gamer} studied how people with physical disabilities streamed for changing disabled viewers' negative mindsets about disabilities. In addition, work by Johnson~\cite{johnson2019inclusion} presented the enjoyable experiences and challenges of video game streamers with physical disabilities, ADHD (Attention Deficit Hyperactivity Disorder), and dyslexia on Twitch. Moreover, some prior work focused on economic and employment opportunities of live-streaming for disabled people ~\cite{johnson2019inclusion, johnson2020disability,qu2020internet}. Recently, Jun et al.~\cite{10.1145/3476038} explored the motivations, practices, and challenges of streamers with visual impairments on Youtube and Twitch. 
Although this work uncovered streamers' motivations (e.g., achieving personal goals, delivering messages, overcoming limitations of recorded videos, getting inspired \rv{by} other streamers) and challenges (e.g.\rv{,} multitasking, screen reader related issues), it did not investigate BLV streamers' experiences with content-curation algorithms adopted by the platforms. However, recent research suggested that users from marginalized groups perceived live-streaming platforms 
may have algorithmic biases against minority \rv{communities}, such as race and ethnicity minorities and the LGBT community~\cite{karizat2021algorithmic}. 
Therefore, in this work, we take a first step to investigate BLV streamers' experiences with content-curation algorithms adopted by live-streaming platforms. 

\subsection{Algorithmic Biases for Marginalized Groups on Video-sharing platforms}
\label{Algorithmic Biases for Marginalized Groups on Video-sharing Platforms}

Content creators of video-sharing platforms were aware that algorithms were a crucial factor, which influenced whether their content could be well-known among passersby ~\cite{bucher2017algorithmic, trewin2019considerations,pedersen2019my,wu2019agent}. Prior research ~\cite{simpson2021you, linessocial,trewin2019considerations} has presented how algorithms excluded the videos created by people from marginalized groups, such as people of color and those who came from LGBTQ groups or lower social-economic class.  For example, several studies indicated algorithmic bias for sexual orientation or other gender issues. Simpson et al.~\cite{simpson2021you} described that \rv{users from the LGBT+ community believed that} the Tiktok algorithms censored or removed the content created by \rv{the LGBT+ group}.  Haimson et al.~\cite{haimson2021tumblr} \rv{reported that users perceived that} the algorithms of Tumblr sometimes oppressed the content from trans through classifying them into adult content. Also, recently, some anecdotes~\cite{linessocial,trewin2019considerations, bozdag2013bias,robertson2019tiktok,biddle2020invisible,biddle2021tiktok,melonio2020tiktok} revealed that in the process of content evaluation, Tiktok limited the popularity of videos created by people with, for example, disabilities, facial disfigurements, Down syndrome, or autism. \rv{Tiktok prevented their} content from reaching non-disabled people's "For you" feed.

The \rv{perceived} algorithmic biases and suppression enabled content creators to come up with some coping strategies. For example, Simpson et al.~\cite{simpson2021you} showed that content creators recapped or reposted the content removed by the Tiktok algorithms to push back against the \rv{perceived} suppression. \rv{According to work by Karizat~\cite{karizat2021algorithmic},} content creators \rv{believed that the Tiktok algorithms }suppressed \rv{ the content related to marginalized }social identities. \rv{Content creators} resisted the Tiktok algorithms through individual actions~(e.g., sharing the content perceived to be suppressed ), collective actions~(e.g., collectively commenting and liking the videos silenced by algorithms), and altering the performance of content.

Despite increasing attention to algorithmic fairness for marginalized groups, whether BLV users perceive their contents are impacted unjustly by the algorithms of live-stream platforms remains unclear, which motivated our work.

\begin{table*}[ht]

\begin{tabular}{l|l|l|l|l|l}
\hline
\textbf{Participant ID} & \textbf{Gender} & \textbf{Vision Condition} & \textbf{Age} & \textbf{Occupation}              & \textbf{Education}  \\ \hline
\textbf{P1}          & Female          & Low vision             & 33-37       & Company Employee                 & Bachelor's Degree   \\ \hline
\textbf{P2}          & Female          & \rv{Totally blind}         & 33-37       & Company Employee                 & Bachelor's Degree   \\ \hline
\textbf{P3}          & Male            & \rv{Totally blind}         & 28-32       & Company Employee                 & Bachelor's Degree   \\ \hline
\textbf{P4}          & Female          & \rv{Totally blind}         & 33-37       & Company Employee                 & Bachelor's Degree   \\ \hline
\textbf{P5}          & Female          & \rv{Totally blind}         & 28-32       & Self-Employed                   & Bachelor's Degree   \\ \hline
\textbf{P6}          & Male            & Low vision             & 28-32       & Masseur                         & High School or Less \\ \hline
\textbf{P7}          & Male            & Low vision             & 28-32       & Masseur                         & Associate's Degree  \\ \hline
\textbf{P8}          & Male            & \rv{Totally blind}         & 33-37       & Masseur                         & High School or Less \\ \hline
\textbf{P9}          & Male            & \rv{Totally blind}         & 23-27       & Company Employee                & Associate's Degree  \\ \hline
\textbf{P10}         & Female          & \rv{Totally blind}         & 28-32       & Unemployed                       & High School or Less \\ \hline
\textbf{P11}         & Male            & Low vision             & 28-32       & Masseur                         & High School or Less \\ \hline
\textbf{P12}         & Male            & \rv{Totally blind}         & 38-42       & Self-Employed                   & High School or Less \\ \hline
\textbf{P13}         & Female          & Low vision             & 18-22       & Masseur                         & High School or Less \\ \hline
\textbf{P14}         & Male            & \rv{Totally blind}         & 33-37       & Masseur                         & High School or Less \\ \hline
\textbf{P15}         & Male            & Low vision             & 33-37       & Masseur                         & Associate's Degree  \\ \hline
\textbf{P16}         & Male            & Low vision             & 28-32       & Company Employee                 & Associate's Degree  \\ \hline
\textbf{P17}         & Male            & \rv{Totally blind}         & 28-32       & Masseur                         & Associate's Degree  \\ \hline
\textbf{P18}         & Female          & \rv{Totally blind}         & 28-32       & Self-Employed                   & High School or Less \\ \hline
\textbf{P19}         & Male            & Low vision             & 28-32       & Masseur                         & Associate's Degree  \\ \hline
\end{tabular}

\small
\centering
\Description[Summary of BLV streamers interviewed]{A table with 5 columns and 20 rows. Participants' Information is given, including ID, gender, eye condition, age, occupation, and education.}

\caption{Summary of BLV streamers interviewed. Among the 19 participants, 7 were female and 12 were male, aged from 19 to 42. 11 were blind and 8 were low-vision. Participants were engaged in four kinds of occupations}
\label{table:participants}
\end{table*}

\section{METHOD}
\label{METHOD}

To better understand BLV streamers' motivations, practices, challenges, and coping strategies,  we conducted \rv{an IRB-approved} semi-structured interview \rv{study} with 19 BLV streamers, who had live-streamed over ten times and over half an hour at least three times in the last three months on Douyin. 

\subsection{Participants}
\label{Participants}
Table 1 shows participants' demographic information, including age, gender, vision condition, and occupation. Eleven~(N=11) participants were totally blind, and eight~(N=8) were low vision. Seven were female, and twelve were male. Participants were between 18 and 42 years old~($M=31, SD=4$). 
\rv{It was worth noting that most of our participants were not tech-savvy, who mainly utilized the screen readers and Voice-over to access live-streaming platforms.} 

Our participants were recruited by sending direct messages on Douyin or sending messages to their WeChat posted in their profiles on Douyin or snowball sampling. We searched on Douyin, using BLV-related keywords~(e.g., blind, visual impairment, promote accessibility, etc.), and sent messages directly to the eligible streamers. We contacted 82 streamers directly in total, and 6 were willing to accept our interview. Through snowball sampling, we recruited another 13 participants. It was worthy to note that the BLV participants we recruited had diverse educational levels, ranging from high school degree or less to bachelor's degree.

We watched and engaged in their live-streaming two weeks before the interviews to build rapport and trust with our participants. We also watched their short videos, ``like'', and left comments on the videos. At the same time, their live-streamings and short videos were recorded and saved.

\subsection{Procedure}
\label{Procedure}

We observed their live-streaming practices~(e.g., topics chosen, hashtags added, interacting approaches, etc.). Meanwhile, we checked their profiles to gain more useful information the streamers disclosed. This process helped us prepare for semi-structured interviews.

We conducted the interviews between June 22 and August 12 in 2021. All interviews lasted approximately 50 minutes to 3 hours via WeChat voice call, and participants were provided with 100 CNY \rv{(about 16 US dollars)} after interviewing.  All interviews were conducted in the participants' native language~(i.e., Mandarin Chinese) and audio recorded.

We aimed to understand participants' perceptions of algorithms employed by living-stream platforms. We first their general understanding of algorithms. We then asked them how algorithms worked on living-stream platforms. We further asked about their perceptions of the factors related to the algorithmic evaluation of content, the perceived algorithm-related challenges they experienced, and their coping strategies.

\subsection{Data Analysis}
\label{Data Analysis}
Our data included the audio recordings of the interviews, recordings of their live-streaming, short videos they posted, and research field notes. We first transcribed the audio recordings verbatim. Then two native Mandarin-speaking authors coded the transcripts independently using an open-coding approach~\cite{corbin2014basics}. In the regular weekly meetings, the two coders discussed the codes (e.g., definitions and example quotes) and resolved conflicts to consolidate the code book. If there were any unresolved conflicts, the rest of the authors joined the discussion to gain a consensus. Afterward, the research team performed affinity diagramming~\cite{hartson2012ux} to group the codes into clusters based on their semantic similarity, and we identified the themes of the clusters. These themes and codes, along with representative quotes, form the structure of our findings, which will be reported in the next section.



\section{RESULTS}
\label{RESULTS}
\begin{figure*}[ht]
    \centering
    \includegraphics[width=\textwidth]{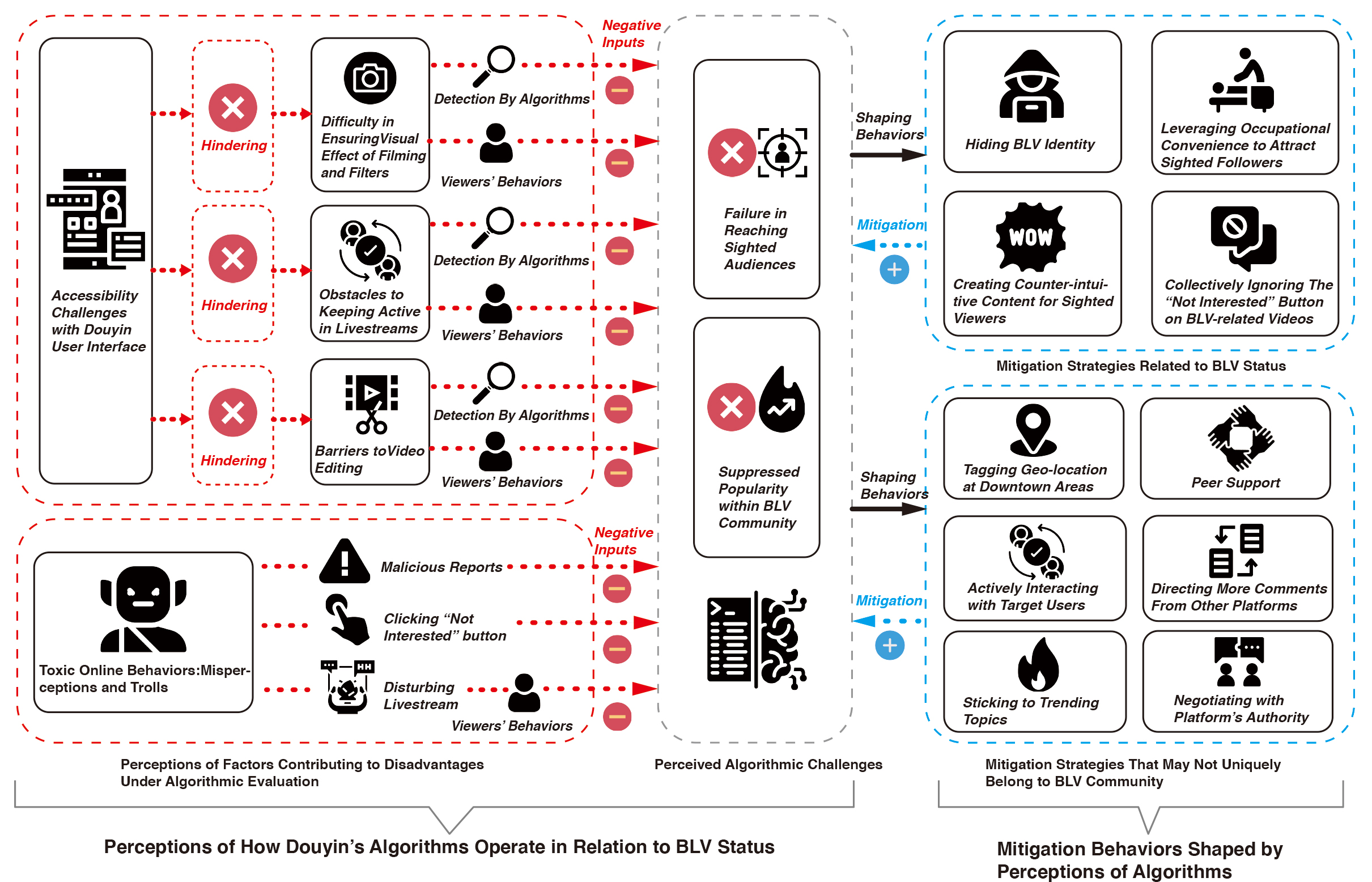} 
    \Description[Schematic of the findings]{Three main columns related to three themes. Under each theme, some bullet points and logic relations were included.}
    \caption{The diagram illustrates BLV streamers' perceptions of how Douyin's algorithms operate in relation to BLV status and their mitigation behaviors shaped by such perceptions. Accessibility challenges with Douyin user interface and toxic online behaviors are perceived factors (i.e., accessibility challenges with Douyin's user interface and toxic online behaviors agains BLV content creators) contributing to negative inputs to Douyin's algorithmic feedback loops, and may exacerbate the perceived algorithmic challenges. The awareness of perceived algorithmic challenges in turn shape users' behaviors to mitigate these barriers. Some mitigation strategies are related to BLV status, others may not uniquely belong to BLV community but demonstrate with affluent details about how such strategies are applied by BLV users.}
    \label{fig:schematic}
\end{figure*}

\rv{The interviewed streamers streamed various contents, such as sharing their daily lives, sharing their professional lives, answering questions about visual impairments, explaining how to use assistive technologies (e.g., screen readers), and socializing (Figure 1) For example, P2 chose to share her daily life with her guide dog through livestreams and short videos to dispel the general public's misunderstanding of guide dogs. One such example was as follows, \textit{``Guide dogs were assistance dogs for people with visual impairments and were well trained. They (guide dogs) would not make any trouble for others.''}} 

Next, we present our findings on \rv{three} themes: 1) \textit{Perceptions of Factors Contributing To Disadvantages Under Algorithmic Evaluation}; 2) \textit{Perceived Algorithmic Suppression};3) \textit{Mitigation Strategies.} \rv{ Figure~\ref{fig:schematic} shows how these themes and sub-themes are connected with each other.}

\subsection{Perceptions of Factors Contributing to Disadvantages Under Algorithmic Evaluation}
\label{Challenges with Douyin Algorithmic suppression and Mitigation Strategies}


Participants perceived the algorithms to own a great deal of power, assessing whether their content could pass the “examination” from Douyin. 
For instance, P17 expressed her perceptions of which factors of a user's content played essential roles in influencing the content evaluation from the algorithms, which was informed by the several articles on new media and word of mouth from Douyin users.
\begin{quote}
   \emph{``There were many metrics in terms of content evaluation, such as the completion rate of your videos, the time users spent on your live room, the number of video views, the `likes' and comments, as well as the virtual gifts you received. It was difficult for us to fit into these rules due to our visual disability.''} (P17)
\end{quote}

Participants felt that accessibility challenges with Douyin's user interfaces resulted in disadvantages under algorithmic evaluation. The accessibility challenges were related to \textit{difficulties in ensuring visual effect filters}, \textit{obstacles to keeping active in livestreams}, and \textit{barriers to video editing}.   

\subsubsection{\textbf{Difficulties in Ensuring Visual Effect of Filming and Filters}}
\label{BLV Content Creators' perceptions of how Douyin's algorithms work}

Participants described that they encountered great difficulty ensuring the visual quality of the content they created. Due to visual impairments, they could not set a proper camera angle that captured the scenes they intended to show. As P2 said, \textit{``I wanted to film the whole body of my guide dog, yet only captured his head in the live-streaming.''} P15 noted that he was often unaware that he had been off the shooting range: \textit{``It was inevitable to move my body during live-streaming, which resulted in only showing my shoulders rather than my upper body in the camera. However, I did not notice it at the moment due to my sight loss.''}

\st{While prior work showed that Instagram users with low vision could apply filters to photos to achieve a specific aesthetic effect, it did not mention the usage of visual filters in live-streaming among BLV streamers. Our work filled this gap and revealed that BLV streamers, especially those who were blind, were unable to apply visual filters to short videos and live-streaming.} Our participants, especially those who were blind, expressed that they were unable to apply visual filters to short videos and live-streaming. The reason for this frustration was they had no idea about the effects that the filters would achieve. P3 explained, \textit{``It seems that visual filters uniquely belong to sighted users. I can not imagine what will happen to my appearance after turning a filter on.''} As a result, image management was also challenging for BLV streamers: \textit{``We could not dress up as sighted people did, and some of \rv{people with visual impairments}'s eyes may not look pretty or even normal to sighted people.''}~(P6)

\textbf{Participants often had difficulties creating visually-engaging content and felt that their content was not valued by Douyin's content promotion algorithms.} They believed that Douyin's algorithms assessed the aesthetic of content by the conventional beauty standards of sighted people. As P15 explained, \textit{``The slogan in Douyin's advertisement reflected the value of Douyin's algorithms: to explore beautiful things. ''}
In addition, P6 described that BLV content creators, especially those who were blind, often could not shoot their faces from a proper position, and felt this could be an important reason why some potential viewers may skip his livestreams.

\begin{quote}
    \emph{``The BLV content creators could not see.~(When using cell phones to shoot,) the camera may capture part of his or her face, even only one eye or the nose…some of BLV content creators were getting too close to the camera that the viewers could only see his or her face. \st{Those viewers may quickly swipe after seeing such images.}~\rv{One of my sighted friends told me, `Several short videos created by people with visual impairments randomly appeared on my For You Page. The visual effects were awful, because I could only see a nose or a forehead rather than a face. Therefore, I quickly skipped those videos.'}''}
\end{quote}
Participants also thought that their shortcomings in managing presented images made their content disadvantaged under the algorithms. As P19 noted, 
\begin{quote}
\emph{``Given that \rv{people with visual impairments}'s visual status, we could not dress up as sighted people did, and we had no idea how to use visual filters and the effects of filters. Some viewers thought we were not good-looking, then left our live room or closed our videos.~\rv{For example, I was randomly paired with a sighted user (by Douyin algorithms), when he noticed my looking, he just said, `Oh, how ugly!' and left the live-stream immediately.}''}
\end{quote}
As a result, participants perceived that the algorithms would think BLV users' content was not worthy of being promoted. As P15 said, \textit{`` Some audiences were not willing to stay and would quickly swipe our videos, not even mention leaving comments or clicking "Like". (We believe)This led to a low completion rate as well as a low score from the algorithms.''}

\subsubsection{\textbf{\st{Interaction with Viewers}~\rv{Obstacles to Keeping Active in Livestreams}}}
\label{BLV Content Creators' perceptions of how Douyin's algorithms work}

 Unlike sighted streamers who can instantaneously view and respond to onscreen text comments, BLV streamers could not immediately know the messages that their viewers typed in the comment area because they needed to suspend the live-streaming and touch the screen to listen to what the viewers said with the help of a screen reader. Also, participants noted that compared to sighted streamers who could skim through comments, it was inevitable for BLV streamers to spend a significant amount of time listening to the entire comments read by the screen reader sequentially. As P9 said, \textit{`` It is inefficient to interact with the audience since the time lag exists in every step of reading comments.''} Additionally, participants expressed that it was challenging to read every comment from viewers during live-streaming because the list of comments kept automatically scrolling in real-time when the audience typed new comments. P6 stated, \textit{``I need to touch the comments one by one. Each time I listen to a comment, I can not touch the following one since the list of comments has scrolled quickly to the bottom for showing the newest messages.''} Participants expressed the concern that such challenges with reading and responding to the comments resulted in the misunderstanding from the audience. As P3 said, \textit{``They may have thought that I intentionally ignored their comments.''}

\textbf{Participants believed that the barrier to actively and timely interacting with the audience resulted in disadvantages under Douyin's algorithmic metric.} Their' comments highlighted beliefs that Douyin's algorithms valued streamers who could keep talking during live-streaming sessions. As P18 said,
\begin{quote}
    \emph{``\rv{We believe that} the algorithms could detect the streamers' voices in the live room. You needed to keep talking or singing in order to make a good impression on the algorithms. Otherwise, the algorithms may oppress your live-streaming.''}
\end{quote}
Such evaluation metrics from the algorithms were perceived as challenging by participants. For example, P18, who streamed to sing songs, found it challenging to keep singing due to the visual impairment: \textit{``It was impossible to sing while looking at the lyrics because I could not see. And I often forgot the lyrics, which induced me to stop singing. Therefore, the algorithms must dislike my live-streaming. ''}
In addition, participants hold the opinions that the algorithms valued those live-streams where streamers actively interacted with viewers because active interaction could contribute to longer viewership duration. However, BLV users could hardly interact with the audience as soon as possible, which resulted in misunderstandings among viewers. As P18 explained,
\begin{quote}
    \emph{``I needed to touch the messages in the comment area to know what happened. Given that I could not keep touching the screen reader during the whole session, it was tricky to say hello to viewers quickly or reply to their comments immediately. \st{They may think I ignored them and left my live-streaming.}~\rv{Sometimes I received complaints from viewers, they made comments like `why was this streamer ignoring audiences? it was impossible not to see these messages!' When I noticed their comments (by touching the screen) and was about to reply, I found they had left.}''}
\end{quote}
This was echoed by P6, who could not show timely appreciation for the audience that gave virtual gifts to his live-streaming. \textit{``They~(the audience) must think I was rude, feeling uncomfortable, and leaving my live room, which led the algorithms to perceive my live-streaming as not attractive.''}

\subsubsection{\textbf{Barriers to Video Editing}}
\label{BLV Content Creators' perceptions of how Douyin's algorithms work}

Participants described that it was difficult to edit the video they shot due to the complex and visually-demanding interfaces of video editing software and the compatibility problems of screen readers: \textit{``the screen readers worked awfully on the video-editing application associated with Douyin...many functions could not be read, such as inserting music, adjusting the volume and locating a particular spot in progress bar...''}~(P14) Moreover, in addition to inaccessible user interfaces of video editing tools~\cite{Seo2020UnderstandingTC}, our participants reported a lack of description of their video's content, which offered them little control over the editing process, such as feeling hesitant about the segments that they should clip out.

\textbf{Participants believed that the inaccessibility of video editing interface led to the difficulties in creating content that was valued by Douyin's algorithms.} Participants felt that the algorithms appreciated the short videos that were well-edited according to Douyin's template. But such video creations relied highly on one's visual ability. For instance, P15, who was low vision, noted that the prevalent videos on the platform were generally following Douyin's templates: \textit{``Those videos were elaborate. In the first three seconds, you had to show affluent information like the template. You also needed to switch the scene precisely according to the rhythm of the background music.''}
P15 went on to express the difficulty of meeting such video editing standards.\textit{``Creating such videos required you to edit through looking at the screen and manipulating the complex interfaces, which was highly visually-demanding. ''}
This perception of the challenge to fit into the Douyin video template was echoed by P17. Basic video editing skills could hardly be performed due to visual disability. For instance, clipping invalid frames from the beginning of short videos.  \textit{``The first three seconds were crucial. Once the video could not show attractive or informative content in the beginning few seconds, viewers may skip my video. As a result, the completion rate would be terrible.''}

In addition to accessibility challenges with Douyin's user interfaces, participants felt that toxic online behaviors against BLV content creators resulted in disadvantages under algorithmic evaluation.

\subsubsection{\textbf{Misperceptions and Trolls}}
\label{BLV Content Creators' perceptions of how Douyin's algorithms work}
Participants shared negative experiences about being trolled due to the misperceptions of viewers: \textit{``They doubted whether I was really blind since they did not know the existence of assistive technology, thinking it was impossible for us to use smartphones, not to mention streaming.''}~(P3) \textbf{Participants took internet trolls as another negative impact on algorithmic evaluation.} They believed that Douyin's algorithms would rate their content as detrimental to the community due to being maliciously reported by trolls: \textit{`` My live-streaming and videos were reported as defrauding by trolls, who thought I pretended to blind in order to obtain money. Being reported must result in ruining algorithms' impression on my content''}~(P13) Participants also \st{assumed}~\rv{expressed} that trolls \st{might} intentionally clicked ``not interested'' for their videos, which contributed to negative input into the algorithms' feedback loop. \rv{P8 noted,``some trolls made comments below my videos, said that `Why their (people with visual impairments) videos still show on my For You Page? I have already clicked `not interested' for many many times! I really do not want to see them angling for sympathy.'''} In some cases, participants noted that some viewers made outrageous comments
to disturb their live-streaming. Such a trolling behavior resulted in losing viewers of \rv{people with visual impairments}'s live-streaming: \textit{``They kept calling me a fraud and said I was consuming the sympathy of Douyin users to attract followers. Other viewers \st{may} feel annoyed with the conflicts of those emotional comments and choose to leave the live room, ~\rv{for instance, one of my viewers said, `I feel so sorry, I have to leave because I cannot stand such a bitter quarrel.'}... (I believe)the algorithms themselves treated this short length of viewership as low attractiveness''}~(P5). 

\subsection{Perceived Algorithmic Suppression}
\label{Challenges with Douyin Algorithmic suppression and Mitigation Strategies}


Participants suffered from two main perceived algorithmic suppression. 

\textbf{Failure in Reaching Sighted Audiences}.
Participants generally believed that their content was mainly ``locked'' within the BLV community by the algorithms and was barely amplified to the sighted audience. For example, P7, a blind massage shop owner who wanted to promote the business through Douyin, noticed that most viewers were BLV users, which did not meet his expectations because he looked forward to attracting sighted customers. As he explained, \textit{``Whenever I observed the video view counts provided by the system, I found few sighted users while most were BLV users~\rv{[based on his experience that BLV viewers' IDs often contain terms relating to `A person with visual impairments' or `a blind person')}. Also, the comments were mainly from \rv{people with visual impairments}.''} The perception of BLV streamers' content not reaching sighted users was echoed by P18, who often chatted with sighted streamers at their live-streaming sessions. As she said, \textit{``When they first met me~(at their live room), most of them felt surprised and said, `wow, you are the first BLV user we encountered these years! We never saw any content created by BLV community members before.'''}  P18 further noted, \textit{``They~(sighted streamers) later told me that they eventually saw the BLV-related content only if they intentionally searched the keywords at Douyin, such as blind streamers.''}

Participants' comments spoke to a belief that the algorithms automatically detected their content and account interactions. The algorithms were perceived to mark the account network from followers or viewers who re-posted, liked, or left comments. As a result, they believed that once the algorithms found users with similar networks and behaviors, those users would be confined and categorized as a group with a dedicated user profile. BLV users, under such circumstances, could hardly send the content to massive sighted viewers. For instance, P14, a grocery store owner who wanted to promote goods to sighted people through Douyin while mostly followed by BLV users, elaborated as follows:
\begin{quote}
    \emph{``My online business plan failed...one of my videos were viewed and reposted by several BLV users, which induced more BLV users to follow my account. Then the algorithms may consistently amplify my content to BLV users. As a result, this process kept looping, which enhanced the algorithms' belief that it made the right decision to recommend my content only within the BLV community.''}
\end{quote}
Participants' remarks expressed a perception that the more BLV-related information an account had, the more likely the account was identified as and clustered into the BLV user group. Hence, BLV users may find it more challenging to promote content to sighted audiences.

\textbf{Suppressed Popularity within BLV Community}.
Participants' comments highlighted a belief that Douyin's algorithms limited the popularity of BLV-related content within the BLV community. For example, P16, a public-spirited blind streamer who aimed at popularizing how to manipulate current assistive technology for BLV users through Douyin, perceived that the algorithms suppressed his live-streaming. As he explained, \textit{``It was last December, the number of my viewers dropped suddenly.''} P19 went on to note that he and other streamers found they could not purchase the Dou+ service, which could help boost the popularity of live-streaming. 
\begin{quote}
    \emph{}{``A pop-up window showed that the themes of our content were not suitable to be boosted. You know, once you could not buy such a service, then the popularity of the account was permanently suppressed.''}
\end{quote}
Other participants expressed that not only their live-streaming but also the short videos they created were not valued by the algorithm. P7, a blind content creator who took videos to promote available assistive technology for \rv{people with visual impairments}, shared an example of how Douyin held up his video. As he said, 
\begin{quote}
    \emph{``I had a video about how to quit the screen readers, I tried to upload it to Douyin several times, but the system just told me I failed to upload the video because the platform was overloaded. However, when I uploaded a non-BLV video, it went through''}
\end{quote}
This perception was echoed by P6, a user with low vision, who observed that compared to sighted users' videos, the window that showed the ``not interested button'' popped up at BLV content creators' videos with higher frequency. As he explained, 
\begin{quote}
    \emph{``One of my BLV classmates also noticed that such an indicative window seldom showed up in sighted users' content while it often popped up on BLV users' videos. The algorithms may not value BLV-related videos and thus provided the~(not interested) button to suppress our videos further.''}
\end{quote}
 Participants' comments about the ``not interested button'' expressed a perception that even for BLV audiences, Douyin's content promotion algorithms may consider the BLV-related videos they viewed to be of low quality and not worthy of being promoted within the BLV community. It is worth noting that participants also perceived that sighted people most likely encountered such a ``not interested button'' when they saw BLV-related content on their For You Page. Participants believed that if sighted viewers clicked the button, the algorithm would likely not recommend their live-streams/videos to sighted populations (as Section 4.1.4 
reported).


\subsection{Mitigation Strategies}
\label{Mitigation Strategies: Reaching Sighted Users}
 Our findings show that BLV participants took various actions to mitigate the perceived algorithmic suppression. Participants adopted the following strategies related to their \textbf{BLV status}:~\textit{Hiding BLV Identity}, \textit{Creating Counter-intuitive Content for Sighted Viewers}, ~\textit{Leveraging Occupational Convenience to Attract Sighted Followers}, and \textit{Collectively Ignoring The ``Not Interested'' Button on BLV-related Videos}. 

\subsubsection{\textbf{Hiding BLV Identity}} Participants noted that the hashtags of videos played a significant role in algorithmic user profile grouping and automatic viewership setting. Thus, participants intentionally avoided using hashtags that include BLV-related information, such as blind people, screen readers, etc. For example, P17, who intended to promote his massage shop to sighted people, shared his strategies. 
\begin{quote}
    \emph{``I used to set the hashtags as blind massage, which attracted many BLV users to view my videos. It would make my situation worse if I kept using this hashtag. Thus, I decided to rename the hashtag as healthy massage, keeping a healthy lifestyle, or osteopath. It worked well!''}
\end{quote}

In addition to not using BLV-related hashtags, some participants chose to hide their BLV identity completely. They created new accounts without showing any BLV-related information because they believed intuitive connections with BLV identity led the algorithms to confine their content within the BLV community. They detailed their strategies: \textit{avoid disclosing BLV identity by user name}, \textit{wearing sunglasses to cover their eyes}, \textit{avoiding following or being followed by BLV users}, and \textit{
never posting BLV-related content}. As P6 described, \textit{``It felt like creating an account of a sighted user.''} P6 further described that he only posted non-BLV-related content, such as what happened around the city he lived in, and kept following nearby sighted users, expressing how this strategy worked. \textit{``Some of them followed me back. When my sighted followers reached around 200, I found that some of my videos were viewed more than thirty thousand times by sighted people!''} P6 went on to share his experience about reassuring his video was promoted to nearby sighted users. \textit{``I encountered some unacquainted sighted people when I hung out. They said, 'wow, I know you, you are that [user name] on Douyin. I often watch your videos.' ''} Participants noted that they used the new accounts and the prior BLV-related accounts simultaneously since they perceived maintaining two accounts as a way to balance the tension of promoting content to sighted users and sustaining BLV identity. As P19 said, \textit{``I do not mean to repel the BLV community since I am a member of it. So I use two accounts, one for following and interacting with BLV users, and another for promoting my content to sighted audiences. ''}

\subsubsection{\textbf{Creating Counter-intuitive Content for Sighted Viewers}} Participants described that it was helpful to promote the counter-intuitive content that challenged Chinese sighted users' stereotype about \rv{people with visual impairments} because most of the Chinese sighted people were unfamiliar with the life of the BLV community. For instance, P7 noted that as long as the BLV users posted some videos about independently completing the tasks that sighted people perceive as impossible for a BLV person, such videos would most likely become prevalent within the sighted user community. As he explained,
\begin{quote}
    \emph{``It was a stereotype of mainstream society that \rv{people with visual impairments} could not independently do many things. I witnessed some BLV users' videos that sighted people widely reposted. Those videos showcased how \rv{people with visual impairments} cooked, made coffee, surfed the internet, or built computers by themselves, which introduced strong contrast for sighted users.''}
\end{quote}

P7 went on to describe how such content was reached to the sighted community. \textit{``Even if the algorithms do not actively recommend such content to sighted people, one sighted user may feel inconceivable from his or her random browse. Then the content will be shared and reposted one by one...eventually become widespread.''}




\subsubsection{\textbf{Leveraging Occupational Convenience to Attract Sighted Followers.}} In addition, some participants described how they leveraged occupational convenience to attract sighted followers. P19, a blind massage shop owner, shared how he promoted his Douyin account to more sighted viewers by providing discounts to customers who were willing to follow and further promote the account. He elaborated on it: \textit{``Customers got a 20\% off for next visit if they successfully recommended more than three new followers.''}  Other BLV streamers employed similar strategies to attract sighted audiences. P6 reported that \textit{``Some BLV masseurs may stream how they provide massage services to sighted clients...combining with tagging geolocation at the nearby downtown. It was effective to attract sighted followers while promoting business.'' }  

\subsubsection{\st{\textbf{Providing Mutual Support within BLV User Community}}} \st{Consistent with prior work, BLV users mutually followed BLV accounts, liked and commented on BLV-related content to resist the perceived algorithmic suppression within the BLV community. It was worth noting that participants also described some mutual support strategies that were not discussed by previous research. Participants with more followers described how they helped the less popular users. For instance, P16, a popular BLV content creator in the BLV Douyin community, combined the video content from himself and other users, and the mixed video would be posted on Douyin. As P16 said, \textit{``As my channel has more subscribers, the mixed content posted on my channel with the same theme could help promote others' accounts.''}  Similarly, P19 noted that he created a WeChat group that included all of the BLV streamers he knew, which could help them promote their content within the group. He explained, \textit{``There were more than four hundred people with visual impairments within the group. I thought it would more or less help attract some attention for those with few followers.''}} 

\subsubsection{\textbf{Collectively Ignoring The ``Not Interested'' Button on BLV-related Videos}} Different from prior work~\cite{simpson2021you}, in which LGBTQ TikTok users used the ``not interested'' feature to block the promoted contents that did not align with their identities, our participants noted that they chose to collectively ignore the pop-up window containing the ``not interested'' button, which frequently appeared on BLV users' short videos. They perceived clicking the ``not interested'' button as a way of unconsciously being the conspirator with the algorithms. As P6 explained, \textit{``The algorithms do not like our content; I can not click that button to enhance its value of suppressing BLV content.''}  P18 stated that avoiding clicking the “not interested” button was a way to support BLV users. As she said, 
\begin{quote}
    \emph{``It was challenging for BLV users to create videos on Douyin. I would absolutely not click that button even though I was not too fond of that video since it may discourage their passion for creating content. Instead, I would view the whole videos if they showed up on my For You Page to boost the completion rate.''}
\end{quote}
As reported in section 4.1.4, participants believed that sighted people who encountered and clicked such a button would prevent their streaming from being recommended to sighted viewers. Therefore, ignoring the ``not interested'' button was also perceived as a manner to \textit{``avoid making a bad situation worse''} (P6) and resist the perceived biases against amplifying their contents to sighted viewers.

\subsubsection{\st{\textbf{Leveraging Trending Topics within BLV Community}}} \st{Participants reported that even if their content was not amplified within the BLV community, it was still practical to spread their content through leveraging trending topics within their community. For example, P16 shared that setting the hashtags as the name of White Cane Safety Day led his video to be viewed by more than 100 thousand times. \textit{``The view counts were surprising. I just used that hashtag and made a video to express my festival greetings for the people with visual impairments.''} P7 noted that videos criticizing how people with visual impairments were treated unfairly in daily life had wider dissemination.} 
\begin{quote}
    \emph{``\st{People with visual impairments are mostly sensitive about the unfairness of the mainstream society, such as a rejection of boarding experience from an airline company because of one's visual impairment.  Other BLV users, feeling strong engagement and compassion, will repost or share the content again and again.}''}
\end{quote}


 Moreover, participants also adopted the following strategies \st{\textit{unrelated to BLV status}}~\rv{that \textit{may not uniquely belong to BLV community}}. ~\rv{Although other types of streamers may apply these tactics to mitigate the misalignments between expectations and algorithmic decisions, our findings provided affluent details about how these strategies worked within the BLV community.}
 
\subsubsection{\textbf{Tagging Geolocation at Downtown Areas that Had More Potential Target Viewers}} Participants perceived setting geolocations at the downtown areas as an effective way to resist perceived content suppression from the algorithms. They perceived that the platform seemed to have independent algorithms for geolocation and user profile based on behaviors and interactions patterns. Thus, BLV users could amplify their videos to the sighted users using geolocation tags. P17 reported that he tagged the geolocation associated with videos at a nearby supermarket, finding it useful, \textit{``The number of viewers increased obviously.''} Similarly, P6 said that he tended to tag the location at a middle school located in the downtown area. As he expressed, 
\textit{``This strategy worked, even though the algorithms generally recommended my videos within the BLV community. Given the algorithms about recommending videos to nearby users mainly relied on geolocation, tagging the location could more or less 'open a small window' to help amplify my content to some sighted people.''}

\subsubsection{\textbf{Actively Interacting with Target Users}} Participants described that they actively participated in sighted users' live-streaming since they believed that attracting sighted followers was a straightforward way to promote their content to sighted audiences. P9 expressed that he often took the initiative to click like, write lots of comments, and send gifts to sighted users' live-streaming, which led to attention from sighted streamers. As she said, 
\begin{quote}
    \emph{``They remembered me because of my activeness. After knowing I was a blind user, they tried to help promote my account by introducing me at their live-streaming, advocating for sighted audiences to follow me, and allowing me to join their live chatting without charging gifts. Some of them even proactively joined my live-streaming for the whole session.''}
\end{quote}

Participants also shared their strategies on how they engaged with experienced sighted streamers' live room to learn Douyin's algorithmic impacts. For example, P18, a blind streamer, reported how she learned to play with Douyin's algorithms from those sighted streamers to sell her craft to sighted viewers. As she said, 
\begin{quote}
    \emph{``I learned a ton. I realized that I needed to keep posting craft-related videos rather than a variety of content...it could persuade the algorithms to tag me as a craft content creator, then precisely promote my live-streaming and short videos for those who felt interested in it.''}
\end{quote}

In some cases, participants indicated that actively participating in ``PK''~(i.e., a live video feature in Douyin that enabled two streamers to interact in real-time) with random sighted streamers could effectively attract sighted followers. For instance, P9 expressed that he found all streamers that the system randomly assigned were sighted users. He further noted how he seized the chance to attract those sighted users to follow him. As he explained, 
\begin{quote}
    \emph{``If the sighted streamers did not end the `PK' after knowing I was blind, it meant that they would like to chat with blind people. They often asked questions about the life of blind people, such as how I used the smartphone...I patiently answered their questions for making a friendly impression, which resulted in being followed by them afterward.''}
\end{quote}

\subsubsection{\textbf{\rv{Sticking to Trending Topics.}}} \rv{Participants reported that even if their content was not amplified within the BLV community, it was still practical to spread their content through leveraging trending topics within their community. For example, P16 shared that setting the hashtags as the name of White Cane Safety Day (which was set in an effort to celebrate the achievement of people with visual impairments) led his video to be viewed by more than 100 thousand times. \textit{``The view counts were surprising. I just used that hashtag and made a video to express my festival greetings for the people with visual impairments.''} P7 noted that videos criticizing how \rv{people with visual impairments} were treated unfairly in daily life had wider dissemination.}
\begin{quote}
    \emph{``\rv{People with visual impairments are mostly sensitive about the unfairness from the mainstream society, such as a rejection of boarding experience from an airline company because of one's visual impairment.  Other BLV users, feeling strong engagement and compassion, will repost or share the content again and again.}''}
\end{quote}

\subsubsection{\textbf{\rv{Peer Support}}} \rv{Consistent with prior work~\cite{karizat2021algorithmic}, BLV users followed each other on the platform, liked and commented on BLV-related content to resist the perceived algorithmic suppression within the BLV community. Moreover, participants also described mutual support strategies that were not discussed by previous research~\cite{karizat2021algorithmic, wu2019agent, simpson2021you}. Participants with more followers described how they helped the less popular BLV streamers. For instance, P16, a popular BLV streamer, combined his video content with other BLV streamers' video content to create mixed videos and posted them on Douyin. As P16 said, \textit{``As my channel has more subscribers, the mixed content posted on my channel with the same theme could help promote others' accounts.''}  Similarly, P19 noted that he created a WeChat group that included all of the BLV streamers he knew to help them promote their content within the group. He explained, \textit{``There were more than four hundred \rv{people with visual impairments} within the group. I thought it would be more or less helpful to attract some attention for those with fewer followers.''}} 

\subsubsection{\textbf{Directing More Comments From other Platforms}} Some participants described that the alternative strategy they employed was to have as many comments as possible. For example, P16 was a blind streamer who often answered accessibility-related questions from \rv{people with visual impairments} on Wechat; To direct more comments to his videos, he chose to transfer the Q\&A from WeChat to the comment area of a particular video. As he said,
\begin{quote}
    \emph{``I hoped people use the comment area under the video as a place for chatting, where they enjoy more interactive chats and contribute more conversations. Moreover, other BLV users interested in the topic will join the conversations and feed more comments. ''}
\end{quote}

\subsubsection{\textbf{Negotiating with Platform's Authority}} Some participants described that they actively negotiated with the Douyin authority, hoping that Douyin could help address the challenges regarding suppressed popularity within the BLV community. P19 expressed how he reached out to Douyin customer support. As he explained, \textit{``I complained to them that I could not purchase the Dou+ service; it was unfair. I asked them whether they could ask algorithms engineers about what happened to my account.''} Similarly, P7 stated that he contacted Douyin customer service and complained: \textit{``The video could not be uploaded for several times, I did not buy the reason of platform overload. It must be held up by your algorithms!''} However, the customer service did not respond to participants' complaints.\newline

\section{DISCUSSION}
\label{DISCUSSION}
While prior work ~\cite{10.1145/3476038} described BLV streamer's motivations (e.g., achieving personal goals, delivering messages, overcoming limitations of recorded videos, getting inspired by other streamers) and accessibility challenges (e.g., multitasking, screen reader related issues), our findings extended this line of research ~\cite{10.1145/3476038} by uncovering factors contributing to disadvantages under algorithmic evaluation, and the perceived challenges with algorithmic suppression and mitigation strategies. Based on the findings, we present the following design considerations for building a more accessible and inclusive live-streaming ecosystem. 

\subsection{Design for Inclusive Computer Vision Technology}
\label{How to Help BLV people Ensure Proper Shooting Angles?}
One common recurring difficulty was adjusting the shooting angles properly to ensure that a BLV streamer's face and body were within the camera frame when streaming due to visual impairment. Current face-tracking technology shows the promise to address this challenge, which could detect and locate objects or faces in nearly real-time~\cite{wagner2009multiple,lei2011real,naik2015robust,ataer2016object} on low-end devices like mobile phones~\cite{chen2009streaming}. However, BLV content creators' streaming behaviors complicated the utility of face-tracking technology. Some BLV streamers did not want to show their eyes and thus always wore sunglasses, which made it challenging to track their face since current face-tracking technologies rely on capturing the whole parts of a human face ~\cite{anwarul2020comprehensive,kaur2020facial}. Also, some BLV participants streamed in nearly total darkness because they were not sensitive to the indoor light conditions due to visual impairments, which deteriorated the accuracy of face-tracking technologies ~\cite{rahman2020face}. We suggest that algorithm designers may train the face recognition model with as many BLV streamers' videos as possible.
Additionally, when applying face-recognition technology in BLV users' live-streaming, it is essential to take into consideration how to avoid leaking private information. For example, the faces of the BLV streamer's family members or the photos of faces~\cite{gurari2019vizwiz} in the streamer's room might be accidentally captured. Prior research indicated that privacy was a core value in system design~\cite{friedman2013value} and \rv{people with visual impairments} were concerned about their privacy when using technologies~\cite{10.5555/3361476.3361478,10.5555/3235866.3235879,10.1145/3234695.3236342}. To protect users' privacy, Zhou et al. developed a technology to blur irrelevant people's faces during live-streaming~\cite{9218980}. However, many questions remain unresolved. For example, what visual content, in addition to faces, do BLV streamers perceive as private? Would failures of blurring different privacy-sensitive content cause equal concerns to BLV streamers?

\subsection{Design for Inclusive Virtual Identity Management}
\label{How to Better Help BLV Streamers Do Image Management without Compromising Viewers' Experiences?}

Some participants with eyeball atrophy or who were not good at dressing up also mentioned that they encountered barriers in image management. They worried that their appearance would leave a negative impression on the audience. Hence, some of them tended to wear sunglasses and even turned off the camera directly during streaming. This raises an interesting question of how to help BLV streamers better manage images without compromising viewers' experiences? A possible solution could be to explore the use of virtual idols on short-video platforms that support live-streaming~\cite{lu2021more,black2012virtual,freeman2020my,yin2018vocaloid}. \rv{A virtual idol is a 2D or 3D animated virtual avatar with the voice of a human, which could be operated by individuals or agencies. An increasing number of streamers already started to use virtual avatars in their live-streams, such as voice actors, amateur and professional musicians, and artists~\cite{lu2021more}. One streamer named CodeMico on Twitch used virtual avatars to achieve visual effects, such as blowing herself up and flying into space in the live-streaming, that would otherwise be impossible to perform by humans~\cite{virtual2022code}. Moreover, live-streams with virtual avatars had a broad audience in China~\cite{bilibili2020vtuber,lu2021more}. By June 2021, over 30,000 active streamers conducted live-streams with virtual avatars on Bilibili and gained over 560 million Danmu (i.e., scrolling comments)~\cite{bilibili2020vtuber}. Recently, some content creators with disabilities also began to use virtual avatars. For example, a content creator with paralysis recorded his videos with virtual avatars rather than his real appearance. He commented that without showing his appearance, he could avoid being questioned for sad-fishing with his disability~\cite{disablity2021virtual}. Thus, we believe that} if streamers or video creators with disabilities do not want to show their faces, they could use virtual avatars. However, it remains an open challenge of how to inform BLV streamers about the visual appearances of, and more importantly, the audience's reactions to, their virtual avatars. Towards this direction, designers should consider providing explicit audio descriptions of the virtual avatars' features such as gender, skin tone, hairstyles, makeup, accessories, clothing, and ability status~\cite{bennett2021s} for BLV streamers. It is also essential to \rv{inform BLV streamers promptly about their viewers' reactions and preferences of their avatars}. 

In addition, when it comes to enabling BLV streamers to create and craft their virtual avatars, designing BLV user-friendly audio instructions and avatar building interfaces remains unclear. Thus, it is imperative for designers to consider lowering the burden of designing avatars while ensuring design quality. Furthermore, given that some BLV content creators preferred not to show their BLV identities in live-streaming or videos, a potential tension between BLV identity and avatar's appearance may emerge, which calls into questions: Is the BLV streamer willing to show his or her BLV identity through a virtual avatar? What are the viewers' reactions to the avatar that discloses BLV identity? How should BLV streamers balance maintaining disability identity and keeping the avatar attractive? It is worth investigating different stakeholders' perceptions of the avatar design of BLV streamers.

\subsection{Design for Inclusive Content Creation Experience}
\label{Design for Inclusive Content Creation Experience}
\textbf{\textit{Accessible Filters Editing Experience.}} \st{As discussed in Section 4.1.1, BLV streamers expressed that filters on Douyin were inaccessible.} ~\rv{While prior work~\cite{10.1145/3173574.3173650} showed that Instagram users with low vision could apply filters to photos to achieve a specific aesthetic effect, it did not mention the usage of visual filters in live-streaming among BLV streamers. Our work filled this gap and revealed that BLV streamers, especially those who were blind, were unable to apply visual filters to short videos and live-streaming.} For instance, some functional buttons of filters were not labeled, such as the percentage of luminance. Participants also emphasized that they could not picture the visual effects the filters may achieve. 
Not only is it essential to ensure all UI buttons have been labeled correctly, but it is also important to provide audio instructions to help BLV streamers develop a prior idea of the functioning of the filters during live streaming.  
However, as people's perceptions of the aesthetic of visual content are personal and subjective~\cite{10.1145/3134756}, it can be challenging for BLV streamers to judge whether a visual filter would meet their expectations. Thus, it is important for screen reader designers to investigate ways to make computer-generated descriptions of filters match their expectations of such filters. 
Additionally, designers of screen readers need to personalize audio descriptions to different kinds of visually impaired users in order to improve intelligibility. For instance, an audio description comprehensible to people with non-congenital visual impairments may be incomprehensible to people with congenital visual impairments. Similarly, designers also need to personalize assistive editing features for streamers with differing levels of vision impairment. Unlike low-vision people, participants with total blindness noted that navigating the filter interfaces and locating the buttons were challenging. One possible solution is to use voice commands or gesture inputs (e.g., tapping, swiping, or even eyelid gestures) for navigation~\cite{Li2020iWink,10.1145/3308558.3314136,Fan2020Eyelid,Li2020iWink,Fan2021Eyelid,Li2017BrailleSketch}. 

\textbf{\textit{Accessible Video Editing Experience.}} Another challenge for assistive editing design is to provide audio descriptions on mobile editing tools (e.g., Jianying, a software that allows users to edit videos on mobile phones and imports videos to Douyin). Although previous works introduced algorithms to generate video descriptions~\cite{Johnson2016DenseCapFC, Krishna2017DenseCaptioningEI,wang2021toward}, such algorithms were designed or trained to recognize videos taken by sighted people. The quality and content of videos taken by \rv{people with visual impairments}, however, may differ from those taken by sighted users. For example, a video shot by a BLV streamer may result in inaccurate or even absurd descriptions for current algorithms due to factors, such as jitters or being out of focus. Therefore, common video quality issues should be carefully considered when designing algorithms to generate audio descriptions of videos shot by BLV streamers. Designers should consider collecting diverse sets of videos shot by BLV streamers in different contexts and training such algorithms on those videos. 

Further, as automatic generated audio descriptions may still contain errors, post-verification is needed. For example, previous research mentioned that \rv{people with visual impairments} tend to ask sighted friends or family members to verify the accuracy of computer-generated descriptions of photo, but this also imposes a burden on those sighted people~\cite{10.1145/2998181.2998364, 10.1145/3134756}. This issue may be exacerbated because of the extra efforts required to evaluate video descriptions due to longer length. Therefore, future work should investigate more effective way to validate the quality of video descriptions.

\subsection{Design For Inclusive Algorithmic Experiences}
\label{How to Design More Inclusive Algorithmic Experiences For BLV Content Creators?}

Our finding regarding BLV content creators’ perceived algorithmic barriers mirrored Karizat et al.'s identity strainer theory~\cite{karizat2021algorithmic}, which expressed that the algorithms acted as a strainer to marginalize a variety of social identities, including ability status. However, in ~\cite{karizat2021algorithmic}, there was only one participant who mentioned his or her observation that Tiktok valued the content of able-bodied streamers but oppressed the content created by people with disabilities. We extended Karizat et al. 's work~\cite{karizat2021algorithmic} by providing an in-depth investigation of BLV content creators' perceived challenges from Douyin's algorithms. Participants believed that they were vulnerable under the evaluation metric of the algorithms due to the harmful trolling behaviors and the lack of visual ability to ensure content’s visual appearance, timely interaction with audiences during streaming, and quality of video editing. 
Participants perceived that content created by BLV content creators was seldom amplified to sighted audiences. Instead, they believed that algorithms limited the popularity of their content within the BLV community. 
 
Participants adopted BLV status-related strategies to mitigate perceived algorithmic biases, including 1) hiding BLV identity; 2) creating counter-intuitive content; 3) leveraging occupational convenience in massage shops to attract sighted followers; ; and 4) collectively ignoring the “Not Interested” button on BLV-related videos. Among these strategies, “hiding BLV identity” may lead to conflicts with BLV identities. For instance, in order to “trick” Douyin's algorithms to promote their streaming contents to more sighted viewers, participants avoided disclosing any BLV-related information in their accounts or shied away from any engagement with any BLV-related content and community. Indeed, this tactic may help them reach a wider audience of sighted people, but it would unfortunately force them to hide themselves in non-BLV algorithmic identities~\cite{doi:10.1177/0263276411424420}. Such a strategy is contrary to those in previous works ~\cite{karizat2021algorithmic,simpson2021you}, in which marginalized users tamed the algorithms to make their algorithmic identities~\cite{doi:10.1177/0263276411424420} and their self-identities further closer together. Future research may explore whether content creators from other marginalized groups may use their own status-related mitigation strategies. It may also be fruitful to explore whether there exist tensions when trying to preserve identities and tame algorithms in order reach a certain targeted audience simultaneously. Researchers could also explore how \rv{people with visual impairments} may resolve these tensions?

Participants also adopted strategies that may not uniquely belong to the BLV community to mitigate perceived algorithmic biases, including: 1) tagging locations at popular places; 2) directing more comments from other platforms; 3) actively interacting with target users; 4)peer support; 5) sticking to trending topics and 6) negotiating with the Douyin authority. Since these mitigation strategies are non-BLV status-related (e.g., tagging location and direct comments), they may be applicable to other marginalized content creators (e.g., people of color, people of LGBTQ community, people with other types of disability, or those who are socioeconomically disadvantaged) who want to promote their contents to a broader community but perceive their contents being “locked” within a confined group.
 
Next, we present design suggestions for creating inclusive algorithmic experiences for BLV content creators.
 
\textbf{Accessible Explanation Interface.} Prior research~\cite{10.1145/3173574.3173860} defined algorithmic experience~(AX) as ``an analytic tool for approaching a user-centered perspective on algorithms, how users perceive them, and how to design better experiences with them”, which included five themes for improving AX in the context of social media: algorithmic profiling transparency, algorithmic profiling management, algorithmic awareness, algorithmic user-control, and selective algorithmic remembering. We suggest that algorithmic profiling management of AX 
may be applied to create better AX for BLV content creators, which enables them to modify and tune their profile created by algorithms~(i.e., how algorithms think about them). For example, BLV content creators, who want to promote their 
live-streams or videos to more sighted audiences, could adjust the factors contributing to failure in reaching sighted audiences in their algorithmic profiles, making such profiles aligned with their expectations. However, this raises an open question of how to effectively and sufficiently communicate complex internal operations of algorithms to \rv{people with visual impairments}. Current research on Human-Centered AI leveraged interactive visualizations to allow stakeholders to explore the data and audit the algorithms synchronously and asynchronously~\cite{10.1145/3290605.3300789,Cheng2021SolicitingSF,Fan2020VisTA,Fan2022HAC}. However, such visually-demanding interfaces are inaccessible to \rv{people with visual impairments}. Therefore, it is worth exploring ways to design accessible explainable AI interfaces for \rv{people with visual impairments}, such as combining the potentials of accessible visualization~\cite{Lundgard2019SociotechnicalCF} and dynamic shape-shifting user interfaces~\cite{10.1145/3173574.3173865}.


\textbf{Accessible Participatory algorithms Design.} Our study unpacked insights into BLV streamers' perceived algorithmic biases against their BLV identity, which can be difficult to be effectively incorporated into algorithmic design~\cite{10.1145/3274463}. Towards this challenge, participatory algorithm design may be a promising direction to explore~\cite{10.1145/3359283}. Participatory algorithm design involves stakeholders building the computational models that represent their motivations, values, and goals. However, it remains an open question of how to leverage such a participatory algorithm design. For example, how
to ensure the whole model training process accessible to BLV participants? How to deal with the different values and opinions of participants with different levels of visual impairments (e.g., blind and low-vision participants)? How to take the social-economical conditions of BLV participants into consideration to improve the diversity of the design of the participatory algorithms?

\textbf{Social Awareness.} While it is necessary to address the algorithmic challenges by leveraging technical solutions, \textbf{socio-technical solutions are also crucial}. For example, the misperceptions and trolling behaviors from sighted audiences may negatively influence the input into the video-sharing platform’s algorithms feedback loop, enabling algorithms to oppress content created by \rv{people with visual impairments}. Policymakers should make continued effort to mitigate the general public’s stigma towards people with disabilities ~\cite{li2021choose}, addressing the knowledge gap ~\cite{li2021choose} about disability from mainstream society, and strengthening the inclusive awareness of the public. It is also imperative for video-sharing platforms~(e.g., Douyin) to create an inclusive online community for content creators with BLV or other disabilities. Platforms may help promote the content regarding mitigating misperceptions about \rv{people with visual impairments} or promoting accessibility knowledge to a broader audience, outlining respect to people with different ability levels as the community rules, combating the toxic trolling behaviors towards BLV streamers. In addition, participants noted a lack of understanding about how algorithms work to influence content within the less-educated BLV content creators or those located in rural areas. Informed by~\cite{10.1145/3173574.3173860} ’s framework for AX, we suggest that in the future, video-sharing platforms~(e.g., Douyin) should strive for improving the algorithmic awareness of BLV content creators who are vulnerable in education or social-economic status, telling them explicitly how algorithms operate and what sorts of user behaviors affect the feedback loop of the algorithms. Furthermore, it is also necessary for algorithm designers to consider and respect the diverse needs of people with different abilities instead of just focusing on the algorithmic performance.

\textbf{Algorithmic Identity.} The perceived suppression from the algorithms shaped participants’ behavior to break the feedback loop~\cite{10.1145/2702123.2702174} to promote their content to their target audiences, for example, by manipulating the input to the Douyin's algorithms individually or collectively. While prior work~\cite{karizat2021algorithmic, wu2019agent} noted that influencing the input of the feedback loop had the potential to tilt the algorithms, we found that \textbf{a tension emerged when the BLV users wanted to promote their content to both the sighted and BVI community because the strategies they employed for reaching these two communities were paradoxical at times}. For example, the strategies that participants utilized to promote content to sighted people, such as hiding BLV identity, increasing view and like count of sighted people, attracting sighted followers, resulting in continuously influencing the input into algorithms, which might gradually enhance the algorithms’ belief that their content should be recommended to more sighted audiences. This strategy worked at the expense of abandoning the opportunities to make their content show at BLV users’ FYP~(for you page). This tension was echoed by P9, noting \textit{ ``You could only choose one side~(sighted or BLV community) under the current rules of the algorithms. ”} The aspect of algorithmic user-control in Alvarado and Waern’s AX framework~\cite{10.1145/3173574.3173860} showed the potential to address this tension, which empowered users to control the algorithms directly. For instance, algorithm designers can design a feature that enables BLV content creators to craft and switch between different algorithmic profiles~(i.e., promoting content to different categories of viewers ) to click a switch button on their profile pages directly and conveniently.

\section{LIMITATIONS AND FUTURE WORK}
\label{LIMITATIONS AND FUTURE WORK}
In this work, we studied the practices and challenges of BLV streamers when using user interfaces to create and stream contents that were affected by both the algorithms of Douyin and their corresponding mitigation strategies in China. Live-streaming platforms, such as Douyin, Kuaishou, and Tiktok, adopt similar content promotion algorithms. Thus, we expect that many of the findings might be applicable to such similar live-streaming platforms. However, different live-streaming platforms also adopt some unique content promotion algorithms that may affect their user experiences. Moreover, the culture and audience differences exist between different platforms. For example, while Douyin mainly has Chinese-speaking audiences, Tiktok has mostly English-speaking audiences. More research is warranted to investigate whether and to what extent the perceived algorithmic biases and other practices and challenges differ between different platforms for BLV streamers. 

Our work uncovered BLV streamer's \textbf{perceived} biases of Douyin content promotion algorithms. Future work should adopt more objective approaches, such as analyzing promotion algorithms themselves or the promoted streams, to study whether such perceived biases objectively exist.~\rv{Moreover, future work should also investigate how BLV content creators audit algorithms (i.e., developing and testing hypotheses about observed problematic algorithmic behaviors), and what the ramifications of algorithmic biases, if any, may be.}   
Furthermore, our study examined the live-streaming practices and challenges of people who are blind or have low vision. It remains unknown whether and to what extent live-streaming platforms are accessible to streamers with other forms of disabilities, such as hearing or motor disabilities, and whether these streamers perceive biases in the content promotion algorithms. To this end, it is imperative to investigate the accessibility of the user interfaces and the experiences of the content promotion algorithms of live-streaming platforms for streamers with different disabilities. Such efforts, along with ours, would collectively help identify issues to be addressed to make live-streaming ecosystems a more inclusive place for people with different disabilities.

\section{CONCLUSION}
\label{CONCLUSION}
In this paper, we have presented a qualitative study of BLV Streamers' perspectives on algorithms of the live-streaming platform, Douyin. We presented the findings from a semi-structured interview with nineteen BLV streamers. 
We identified the BLV content creators' perceptions of factors (i.e., challenges with user interface and toxic online behaviors against BLV content creators) that have negative impacts on algorithmic evaluation of their content. We then identified the perceived algorithmic barriers of BLV users as content created by people \rv{with visual impairments} was not amplified to sighted audiences, and the popularity of BLV related content was limited within the BLV community. Followed by the perceived algorithmic challenges, we identified the mitigation strategies people \rv{with visual impairments} employed to make their content reach target audiences, sighted people~(i.e., hiding BLV identities, tagging geolocation at downtown area, creating counter-intuitive content, leveraging BLV-related occupational convenience and actively interacting with sighted users) or BLV users~(i.e., providing mutual support within the BLV community, collectively ignoring the ``not interested'' button on BLV-related video, directing more comments from other platforms, leveraging the trending topic within the BLV community and negotiating with Douyin authority). These findings contributed to our understanding of how BLV streamers perceived they were marginalized and excluded by the Douyin ecosystem, and informed the design considerations for future research in designing a more accessible Douyin interface, and creating more inclusive and equitable algorithm experience.

\bibliographystyle{ACM-Reference-Format}
\bibliography{main}

\end{document}